# Modeling and Experiment of the Suspended Seismometer Concept for Attenuating the Contribution of Tilt Motion in Horizontal Measurements

[1,2] F. Matichard, [1] M. Evans, [1] R. Mittleman, [1] M. McInnis, [1,2] S. Biscans, [3] K.L. Dooley, [1] H. Sohier, [1] A. Lauriero, [1] H. Paris, [1] J. Koch, [1] P. Knothe, [1] A. Carbajo, [1] C. Dufort

[1] LIGO Laboratory, Massachusetts Institute of Technology, Cambridge, MA
[2] LIGO Laboratory, California Institute of Technology, CA
[3] University of Mississippi, Oxford, MS

## Abstract

Tilt-horizontal coupling in inertial sensors limits the performance of active isolation systems such as those used in gravitational wave detectors. Inertial rotation sensors can be used to subtract the tilt component from the signal produced by horizontal inertial sensors, but such techniques are often limited by the sensor noise of the tilt measurement. A different approach is to mechanically filter the tilt transmitted to the horizontal inertial sensor, as discussed in this article. This technique does not require an auxiliary rotation sensor, and can produce a lower noise measurement. The concept investigated uses a mechanical suspension to isolate the inertial sensor from input tilt. Modeling and simulations show that such a configuration can be used to adequately attenuate the tilt transmitted to the instrument, while maintaining translation sensitivity in the frequency band of interest. The analysis is supported by experimental results showing that this approach is a viable solution to overcome the tilt problem in the field of active inertial isolation.

## 1    Introduction

Inertial sensors used to measure horizontal acceleration are also sensitive to tilt motion due to the component of gravity along the axis of the instrument. This is illustrated in Fig. 1, using a passive seismometer (geophone) as an example of an inertial sensor.

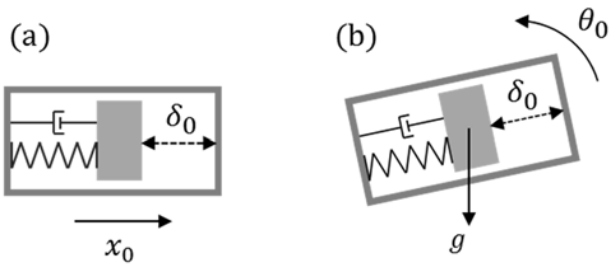

**Fig. 1.  Horizontal inertial sensor response $\delta_0$ to translation $x_0$ (a), and tilt $\theta_0$ (b).**

The proof mass of the instrument is mounted on a spring-damper, and the relative motion between the proof mass and the case is measured to produce the inertial measurement $\delta_0$. Seismometers typically measure the derivative of $\delta_0$, but for the purpose of this discussion it is simpler and equivalent to analyze $\delta_0$. In Fig. 1 (a), the input translation ($x_0$) produces the inertial signal ($\delta_0$). When subjected to tilt ($\theta_0$), the gravitational force along the sensing axis also produces inertial signal ($\delta_0$), as shown in Fig. 1 (b).

Using a small angle approximation, the internal motion can be written as a function of the input acceleration and tilt motion as shown in Eq. (1). The Laplace variable is $s$ and $H$ is the response of the seismometer given in Eq. (2) as a function of the natural frequency ($\omega$) and the damping ratio ($\mu$).

$$\delta_0 = H\,(s^2\, x_0 + g\, \theta_0) \quad (1)$$

$$H = \frac{1}{s^2 + 2\mu\omega s + \omega^2} \quad (2)$$

In the following sections, we will often analyze and plot the inertial motion normalized by the sensor response ($\hat{\delta}_0$), as shown in Eq. (3). This form permits us to analyze the relative contribution of the translation and tilt input motions independent of the response of a specific instrument.

$$\hat{\delta}_0 = \frac{\delta_0}{s^2 H} = x_0 + \theta_0\, \frac{g}{s^2} \quad (3)$$

We define the Tilt Horizontal Ratio (THR) in Eq (4). It is the ratio of the instrument response induced by tilt to the instrument response induced by translation. Subscript 0 is used to denote the THR of a standard seismometer and subscript 1 will be used for the suspended seismometer presented in the following sections.

$$THR_0 = \frac{\partial \delta_0}{\partial \theta_0} / \frac{\partial \delta_0}{\partial x_0} \qquad (4)$$

Equations (1) and (4) are combined to solve for $THR_0$ as shown in Eq. (5). It is a function of gravity and frequency. Due to the $\frac{1}{s^2}$ dependence, measurements tend to be dominated by tilt at low frequency, and by translation at high frequency. The frequency at which they have equal contribution depends on the ratio of the input translation and input tilt spectrums, and is typically between 30 and 300 mHz.

$$THR_0 = \frac{g}{s^2} \qquad (5)$$

The dual sensitivity of inertial sensors to tilt and translation has been a recurrent issue in seismological studies [1-6]. It is also problematic for active isolation systems which rely on inertial sensors [7, 8]. Tilt coupling limits the performance of the seismic isolation platforms used in gravitational-wave detectors, as the signal of horizontal inertial sensors used for active isolation is dominated by tilt at low frequency [9].

Various techniques can be used to reduce the cross-coupling between the translational and rotational degrees of freedom of isolation systems [10, 11], but they have no effect on the tilt induced by ground rotation [12].

Inertial rotation sensors (or tilt estimates based on a combination of two vertical inertial sensors) can be used to measure the tilt motion and subtract it from the horizontal measurement [13-16]. Such techniques are often limited by sensor noise [12]. Only very sensitive rotation sensors can reach the sensor noise tolerable for improving the current performance of active isolation systems used in gravitational wave detectors [17-20].

Passive pendulum filters, often called suspensions, can be used both to isolate the components of a sensitive experiment from ground motion [21, 22], and to build inertial sensors [23-25]. In this article, we investigate the use of suspensions to filter the transmission of tilt motion to horizontal inertial sensors [12]. Not only does this approach not require a rotation sensor to estimate tilt and subtract it from the horizontal measurement, but it can also achieve higher sensitivity as discussed in the following sections.

In the next sections, the passive seismometer is used as a model inertial sensor for our theoretical development. The conclusions are valid for other types of inertial sensors such as force-feedback broadband seismometers and accelerometers. Gravity is used as a reference for the definition of the inertial frame. We assume that gravity is constant over the measurement period. We acknowledge that this assumption may not hold at very low frequency, but it is a reasonable assumption in the frequency band of interest discussed in this article (i.e., greater than 1 mHz).

The second section of this article provides a model of the suspended seismometer concept, and analyses the tilt filtering obtained with an ideal suspension. It discusses one of the main limitations of this approach which is caused by the related filtering of the translational motion. The third section presents experimental results and shows the attenuation of the tilt signal obtained with the suspension. The fourth section presents results of translation measurements and discusses the loss of translational sensitivity. The fifth section summarizes lessons learned during the prototyping phases, and discusses prospects for the use of such instruments in gravitational-wave detectors.

## 2 The Suspended Seismometer Concept

### 2.1 Motivations and limitations

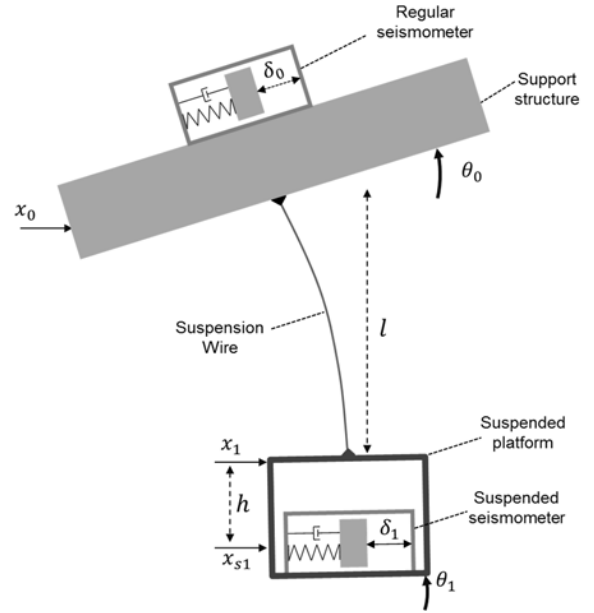

**Fig. 2.** Suspended seismometer concept

The principle of the suspended seismometer concept is illustrated in Fig. 2. The goal is to measure the input translation $x_0$ with minimum contribution of tilt $\theta_0$. A seismometer regularly mounted (top of Fig. 2) measures $\delta_0$ with the ratio of contributions as defined in Eq. (5). To measure the input translation with less contribution from tilt, a seismometer is mounted on a suspended platform which filters the translation and rotation transmission in a way that reduces the THR. The support structure can be either rigidly connected to the ground, mounted on passive components to provide passive isolation, or it can be a stage of an active isolation system. The suspension, which is designed to filter the tilt transmission, also filters the translation transmission. The properties of the suspension must be appropriately chosen to minimize the tilt transmission, while maximizing the response to input translation. The following sections analyze the theoretical response of the suspended seismometer ($\delta_1$), defined in Eq. (6), and its tilt horizontal ratio ($THR_1$), defined in Eq. (7), where $x_{s1}$ is the translation at the sensor location and $\theta_1$ is the tilt of the suspended platform.

$$\delta_1 = H\,(s^2 x_{s1} + g\theta_1) \tag{6}$$

$$THR_1 = \frac{\partial \delta_1}{\partial \theta_0} \Big/ \frac{\partial \delta_1}{\partial x_0} \tag{7}$$

### 2.2 Equations of motion

Fig. 3 shows the forces and motions at both ends of the suspension wires, using subscript 0 for the top end and subscript 1 for the bottom end. The forces are labelled $F$, the torques $\tau$, the translations $x$, and the rotations $\theta$. The wires are under tension (mass $m$ and gravity $g$).

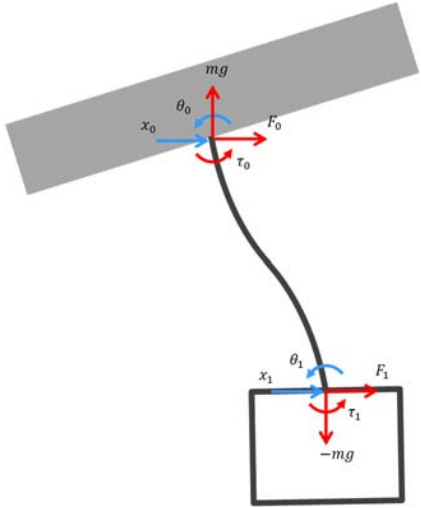

**Fig. 3.** Degrees of freedom and forces on the wires.

The forces at the joints between the wires and the suspended platform can be expressed as a function of the imposed motion ($x_0, \theta_0$) and the degrees of freedom ($x_1, \theta_1$) as shown in Eq. (8) using the stiffness matrices $\boldsymbol{K_0}$ and $\boldsymbol{K_1}$ given in Eq. (9) and (10). The components of these matrices given in Eq. (11) to (14) are a function of the wire length $l$, the wire's moment of inertia $I_a$, the tension ($mg$), and the Young modulus $E$ which are combined within the parameter $K$ defined in Eq. (15).

When $K$ tends toward infinity, the wire acts as an ideal pendulum link with no stiffness in the joints (see [26] for more details). Consequently, the larger $K$ is, the lower the tilt transmission.

$$\begin{bmatrix} F_1 \\ \tau_1 \end{bmatrix} = \boldsymbol{K_0} \begin{bmatrix} x_0 \\ \theta_0 \end{bmatrix} + \boldsymbol{K_1} \begin{bmatrix} x_1 \\ \theta_1 \end{bmatrix} \tag{8}$$

$$\boldsymbol{K_0} = \begin{bmatrix} k_a & -k_b \\ k_b & k_d \end{bmatrix} \tag{9}$$

$$\boldsymbol{K_1} = \begin{bmatrix} k_a & -k_b \\ -k_b & k_c \end{bmatrix} \tag{10}$$

$$k_a = \frac{mg}{2} \frac{K}{\frac{Kl}{2} - \tanh\frac{Kl}{2}} \tag{11}$$

$$k_b = \frac{mg}{2} \frac{\tanh\frac{Kl}{2}}{\frac{Kl}{2} - \tanh\frac{Kl}{2}} \tag{12}$$

$$k_c = \frac{mg}{2K} \left( \frac{\frac{Kl}{2}\tanh\frac{Kl}{2}}{\frac{Kl}{2} - \tanh\frac{Kl}{2}} + \coth\frac{Kl}{2} \right) \tag{13}$$

$$k_d = \frac{mg}{2K} \left( \frac{\frac{Kl}{2}\tanh\frac{Kl}{2}}{\frac{Kl}{2} - \tanh\frac{Kl}{2}} - \coth\frac{KL}{2} \right) \tag{14}$$

$$K = \sqrt{\frac{mg}{E\,I_a}} \tag{15}$$

Fig. 4 shows the forces on the suspended platform. The mass of the instrument (not shown) is included in the platform, and we assume that the coupling between the moving mass in the seismometer and the suspended platform is negligible.

Fig. 4 also introduces an important parameter called the $d$ value, which is the distance between center of gravity of the platform, and the bottom suspension joint. As discussed in the next section, the $d$ value is a key parameter in tuning the response of the platform.

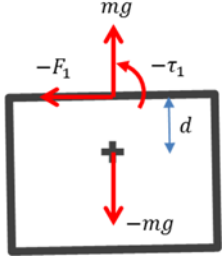

**Fig. 4.** Forces on the suspended platform

The dynamic equilibrium of the suspended platform can be written as a function of the forces at the bottom joint as shown in Eq. (16), where $M_p$ is the inertia matrix given in Eq. (17) and $K_p$ is the stiffness matrix related to the restoring force of gravity given in Eq. (18).

$$\begin{bmatrix} F_1 \\ \tau_1 \end{bmatrix} = -M_p \begin{bmatrix} \ddot{x}_1 \\ \ddot{\theta}_1 \end{bmatrix} - K_p \begin{bmatrix} x_1 \\ \theta_1 \end{bmatrix} \quad (16)$$

$$M_p = \begin{bmatrix} m & md \\ md & I + md^2 \end{bmatrix} \quad (17)$$

$$K_p = \begin{bmatrix} 0 & 0 \\ 0 & mgd \end{bmatrix} \quad (18)$$

The equilibrium of the wires given in Eq. (8) is combined with the platform's equilibrium in Eq. (16) to produce the equations of motion. They are written in the Laplace domain in Eq. (19).

$$\begin{bmatrix} x_1 \\ \theta_1 \end{bmatrix} = -[M_p s^2 + K_p + K_1]^{-1} K_0 \begin{bmatrix} x_0 \\ \theta_0 \end{bmatrix} \quad (19)$$

Before discussing simulation results, the quasi-static response can be analyzed to provide some physical insight. Neglecting the inertial term and assuming $d$ is small (as will be discussed in the next sections), the platform response reduces to Eq. (20) and (21), in which $\lambda$ is the static coupling between input tilt and output translation, and $\varepsilon$ is the residual tilt transmission.

$$\begin{bmatrix} x_1 \\ \theta_1 \end{bmatrix} \sim -K_1^{-1} K_0 \begin{bmatrix} x_0 \\ \theta_0 \end{bmatrix} \quad (20)$$

$$\begin{bmatrix} x_1 \\ \theta_1 \end{bmatrix} \sim \begin{bmatrix} 1 & \lambda \\ 0 & \varepsilon \end{bmatrix} \begin{bmatrix} x_0 \\ \theta_0 \end{bmatrix} \quad (21)$$

This form can be used to approximate the tilt horizontal ratio of the suspended instrument ($THR_1$) at very low frequency as shown in Eq. (22). Both terms are related to the stiffness of the wire. (This formulation assumes that the distance between the instrument and the bottom suspension joint is negligible, which is a good approximation, as will be shown section 2.3).

$$THR_1 \sim \lambda + \varepsilon \frac{g}{s^2} \quad (22)$$

The first term ($\lambda$) can be approximated as shown in Eq. (23), and it can be interpreted as a small portion of rigid wire between the actual top joint location and the effective point of rotation. For a suspension point rotation of $\theta_0$, the platform will rotate by an angle $\theta_1 = \varepsilon \theta_0$. The smaller the stiffness of the wire and the higher the tension, the smaller the two terms of $THR_1$ as shown in Eq. (23)-(24).

$$\lambda \sim \frac{K_b}{K_a} \sim \frac{1}{K} \sim \sqrt{\frac{EI_a}{mg}} \quad (23)$$

$$\lim_{K \to \infty} \varepsilon = 0 \quad (24)$$

In the next section, the equations of motions in Eq. (19) are used to simulate the response of the suspended platform over the bandwidth of interest.

### 2.3 Simulation and analysis

This section uses simulations to analyze the response of the suspended seismometer to input translation, to input rotation, and the tilt horizontal ratio of the suspended seismometer. The parameters used in the simulation are shown in Table 1. They correspond to the values of the experiment presented in the next section, which uses two wires in parallel (see section 5 for the discussion on wires and suspension configurations).

**Table 1**

| Symbol | Name | Value |
|---|---|---|
| $l$ | Wire length | 0.438 m |
| $I_a$ | Wire moment of inertia | 9.89x10$^{-15}$ m$^4$ |
| $E$ | Wire Young modulus | 2x10$^{11}$ N/m$^2$ |
| $m$ | Platform mass | 60 kg |
| $I$ | Platform inertia | 8 kg.m$^2$ |
| $h$ | Instrument location | 0.37 m |

The platform response to input translation ($x_0$) is shown in Fig. 5, assuming the center of mass of the platform is perfectly aligned with the suspension point ($d = 0$), and using a structural damping value factor of 1% (arbitrary value used to introduce damping in the simulations).

The solid black curve shows the translation ($x_{s1}/x_0$) of the platform at the sensor location ($x_{s1}$ is shown in Fig. 2). The transfer function is similar to the response of a point mass pendulum, with a natural frequency that can be approximated as given in Eq. (25).

$$f_p \sim \frac{1}{2\pi}\sqrt{\frac{g}{l}} \quad (25)$$

This simulation shows that the distance $h$ between the bottom suspension joint and the instrument has little influence on the response as written in Eq. (26).

$$\frac{x_{s1}}{x_0} = \frac{(x_1 + h\,\theta_1)}{x_0} \sim \frac{x_1}{x_0} \quad (26)$$

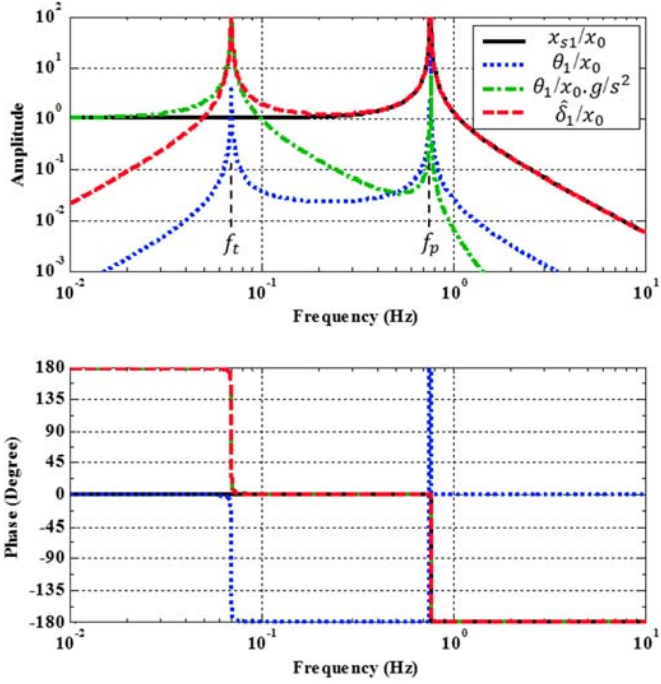

**Fig. 5. Platform and instrument response to input translation ($x_0$)**

The platform output rotation ($\theta_1/x_0$) is shown with the dotted blue line, which has a low frequency resonance, called the tilt mode. This frequency can be approximated by Eq. (27), where $n$ is the number of wires used in parallel to suspend the platform.

$$f_t \sim \frac{1}{2\pi}\sqrt{\frac{nk_c + mgd}{I}} \quad (27)$$

The smaller the $d$ value, the smaller the restoring force of gravity, and thus the smaller the tilt frequency. The $d$ value can be either positive or negative depending on whether the center of mass is positioned above or below the joint. For negative values gravity acts as an anti-spring, such that the platform becomes unstable when the anti-spring cancels out the rotational stiffness of the wires.

The response of the suspended seismometer to translation $x_0$ is the sum of the translation and rotation contributions as shown in Eq.(6). In Eq. (28), the response ($\delta_1$) is normalized by the dynamics of the instrument ($s^2 H$) to perform a generic analysis independent of the response of a particular instrument. The normalized response, $\hat{\delta}_1$ is given in Eq. (29). The first term on the right hand side of this equation is the translation contribution, and the second term is the rotation contribution.

$$\hat{\delta}_1 = \frac{\delta_1}{s^2 H} \quad (28)$$

$$\hat{\delta}_1 = x_{s1} + \theta_1 \frac{g}{s^2} \quad (29)$$

In Fig. 5, the dash-dotted green line shows the tilt contribution to the normalized suspended seismometer response; notice that while the magnitude of this translation to tilt coupling matches the direct translation to translation coupling, it has the opposite sign. The dashed red line shows the sum of contributions from translation and tilt. This curve shows that between the tilt and pendulum frequencies, the translation sensitivity of the suspended seismometer is about unity. Outside of this bandwidth, the response is band pass filtered. A low tilt frequency, and a high pendulum frequency are therefore necessary to maximize the bandwidth of the translation response. The effect of the translation filtering on measurement noise is discussed in the next section.

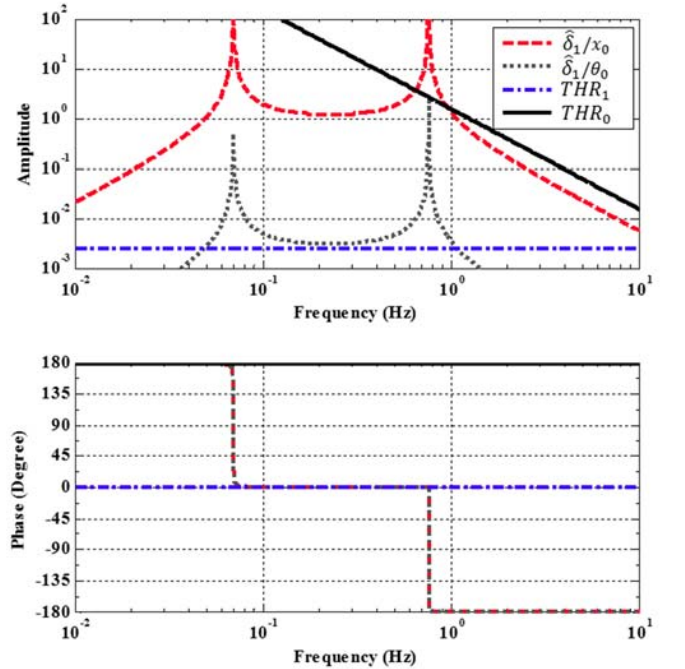

**Fig. 6. Suspended instrument response to translation input $\hat{\delta}_1(x_0)$ and to rotation input $\hat{\delta}_1(\theta_0)$. This plot also shows a comparison of the tilt horizontal ratio of the suspended seismometer ($THR_1 \sim \lambda$) and the regularly mounted seismometer ($THR_0 \sim g/s^2$).**

The response of the suspended seismometer to input tilt is also a band-pass filter. Fig. 6 compares the response to input translation ($\hat{\delta}_1/x_0$) shown with the dashed red line and the response to input tilt ($\hat{\delta}_1/\theta_0$) shown with the dotted grey line. The ratio of these two transfer functions is the tilt horizontal ratio $THR_1$ defined in Eq. (7). It is a constant value, which is approximately equal to $\lambda$, as expected based on the quasi-static analysis given in section 2.2. This tilt horizontal ratio ($THR_1$), shown with the dash-dotted line, can be compared to the tilt horizontal ratio of the non-suspended instrument shown with the solid black line ($THR_0$). For frequencies at which tilt is typically an issue (below $0.1\ Hz$) the tilt horizontal ratio is attenuated by more than four orders of magnitude. Practical limitations of this approach are discussed in section 3 to 5.

### 2.4 Instrument Noise

While the mechanical filtering performed by the suspension greatly reduces the tilt-horizontal ratio, it also reduces the translational sensitivity. To calculate the instrument noise, the self-noise of the instrument is scaled by the response to translation of the suspended seismometer. An example is shown in Fig. 7, using a model of the self-noise of a broadband force feedback seismometer, which is shown with the black solid line.

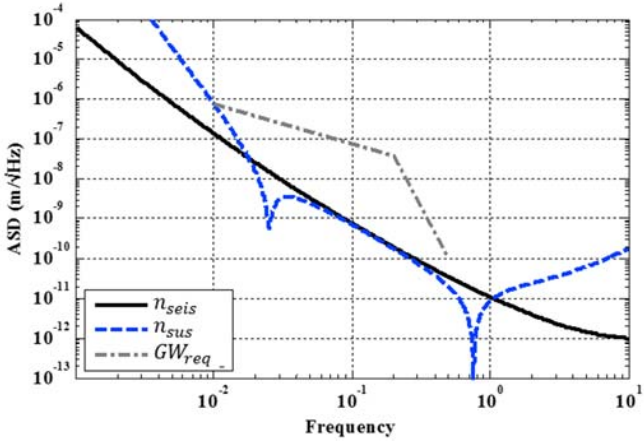

**Fig. 7. Sensor noise of a broadband seismometer ($n_{seis}$), a suspended seismometer ($n_{sus}$), and requirements ($GW_{req}$) for improving the performance of active platforms used in gravitational wave detectors.**

The blue dashed line shows the noise accounting for the filtering induced by the suspension. In this simulation, we assume that the tilt frequency is tuned to 25 mHz. The dash-dotted grey line shows the requirements for improving the performance of active platforms used in gravitational wave detectors [17]. The plot shows that a suspended broadband seismometer tuned with a 25 mHz tilt frequency can significantly improve the performance of the platforms used in gravitational wave detectors at all frequencies of interest.

## 3 Tilt filtering experiment

The simulations presented in the previous section are based on an ideal model of the suspension mechanism. In practice, there are numerous factors that can affect the filtering level, including cross couplings between the degrees of freedom of the platform, misalignments between the suspension axis and the sensing axis, viscous and friction couplings, and unwanted force path between the support structure and the suspended frame (notably induced by the electrical wires of the inertial sensor mounted on the suspended platform).

A suspension prototype and tilt injection platform have been designed to validate the simulation results and quantify the un-modelled residual tilt transmission. A schematic of the experiment is shown in Fig. 8 (top). The support structure (shown in black), is rigidly connected to a large granite table. A voice-coil actuator (force *F*) is used to drive a rotating stage designed to inject tilt at the suspension point. Two stainless steel wires mounted in parallel are used to support the suspension and filter the transmission of tilt from $\theta_0$ to $\theta_1$ (a detailed discussion on the wire and suspension configuration is given in section 5). The top suspension point is aligned with the bearing axis of the rotation stage. The suspended platform shown in blue carries the inertial sensor. A 1 Hz passive seismometer was used to conduct this experiment (results obtained with a broadband seismometer are presented in section 5). Another seismometer is located on the rotating stage to measure the input tilt.

Fig. 8 (bottom) shows the experimental setup. The experiment is performed within a thermal enclosure to reduce the flow of air on the suspended platform. Masses are bolted on top and bottom of the suspended platform to provide the overall mass and inertia specified in Table 1. During the tuning process, small masses are iteratively moved in between the top and bottom plate of the platform to change the *d* value until reaching the desired tilt frequency.

The transfer functions in Fig. 9 show the response of the instrument to the rotation drive. The seismometers signals are calibrated with the theoretical sensor transfer function to plot the normalized inertial motion ($\hat{\delta}_0$ and $\hat{\delta}_1$). In the frequency band of interest (below 1 Hz), the transfer function of the non-suspended sensor ($\hat{\delta}_0/\theta_0$), shown with the black solid line, is in agreement with the expected response ($g/s^2$). The four other curves show the response of the suspended seismometer tuned for four different values of tilt frequency.

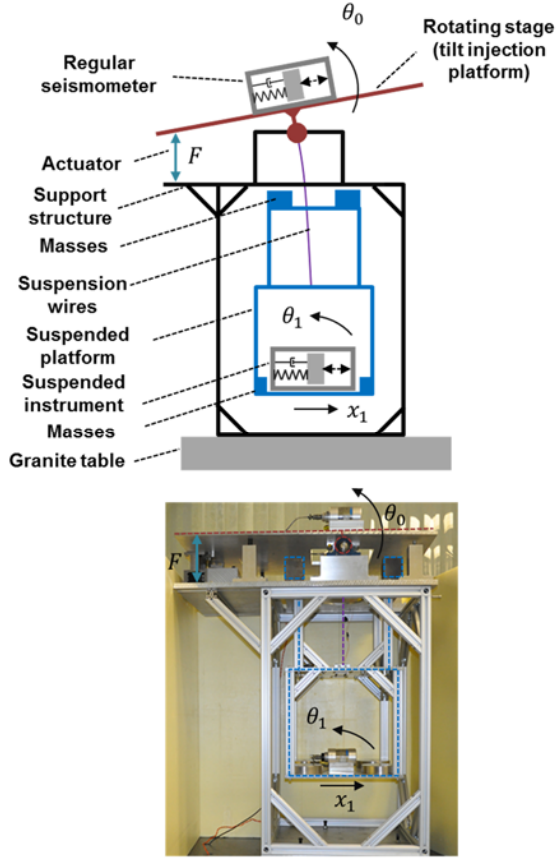

**Fig. 8.** Schematic and actual picture of the experiment designed to inject tilt at the suspension point.

As expected, the pendulum frequency at 0.7 Hz, is not affected by the tuning of the tilt frequency. Above the pendulum frequency the suspension acts as a second order filter. At 3 Hz and above, the transfer functions show several resonances, but this is far beyond the bandwidth of interest.

Below the pendulum frequency, the transfer function of the suspended seismometer is orders of magnitudes lower than the transfer function of the non-suspended seismometer. Below the tilt frequency, the residual tilt coupling of the suspended seismometer is, however, higher than predicted by the ideal model. The lower the tilt frequency of the suspended platform, the higher the extra coupling. The attenuation is still substantial, with values ranging from three to four orders of magnitude depending on the tilt frequency of the suspension.

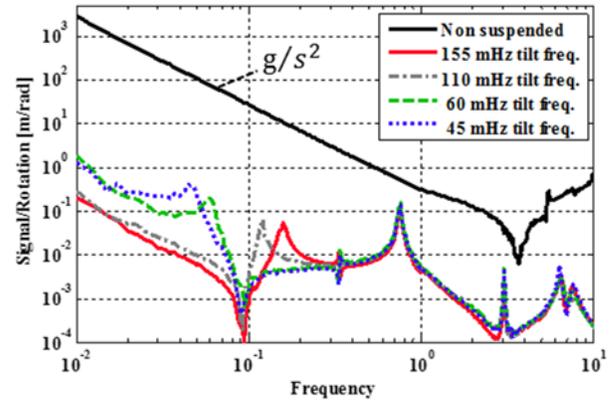

**Fig. 9.** Experimental measurement of the tilt sensitivity of a regularly mounted seismometer ($\hat{\delta}_0/\theta_0$, black solid line) and a suspended seismometer ($\hat{\delta}_1/\theta_0$, other lines)

In Fig. 10, the response of the suspended seismometer tuned for a 150 mHz tilt frequency ("Expe" in the legend) is compared with the response of the corresponding model.

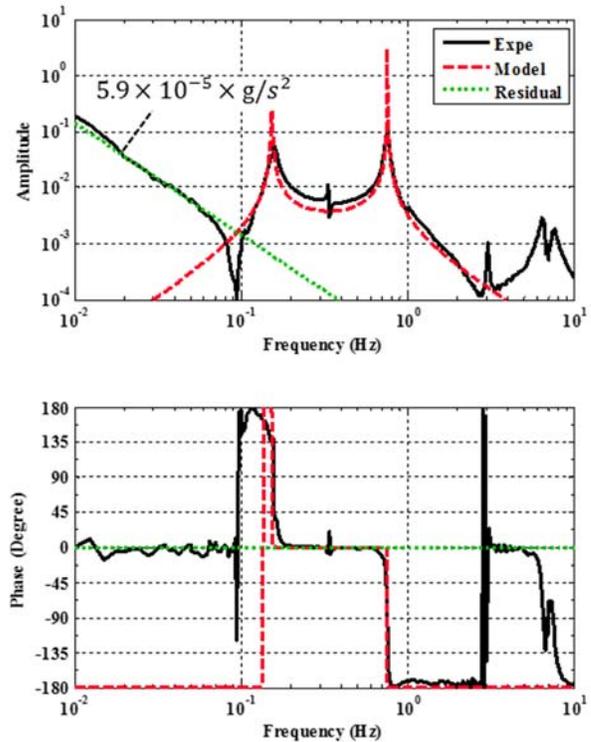

**Fig. 10.** Transfer functions for a tilt drive ($\hat{\delta}_1/\theta_0$). Comparison of experimental results ("Expe" in the legend) and model. The difference (residual coupling) is fitted with the dotted line.

The model does not include the internal dissipation in the passive geophone, which couples with the platform and damp the tilt mode. Consequently, the modes are more damped in the experiment than in the model, but the curves are in good agreement down to 100 mHz. At lower frequencies, the unwanted (and un-modelled) tilt transmission starts to dominate the experiment. The difference between the model and the experiment at low frequency, called residual tilt coupling, is shown with the dotted green line. The coupling value is $5.9 \times 10^{-5}$ $g/s^2$.

A similar plot is shown in Fig. 11, while the tilt frequency of the suspension is tuned at a 45 mHz. In this case, the residual coupling is $3.8 \times 10^{-4}$ $g/s^2$. Sources of cross couplings and residual tilt transmission are discussed in section 5.

This section presented experimental results demonstrating that the suspended seismometer concept is a very effective technique to filter the transmission of tilt motion. Such a reduction in tilt-horizontal coupling could significantly help in improving the performance of the active isolation systems used in gravitational wave detectors.

The next section presents tests and analysis of the main drawback of this approach, which is the reduction of the translation sensitivity induced by the suspension filtering.

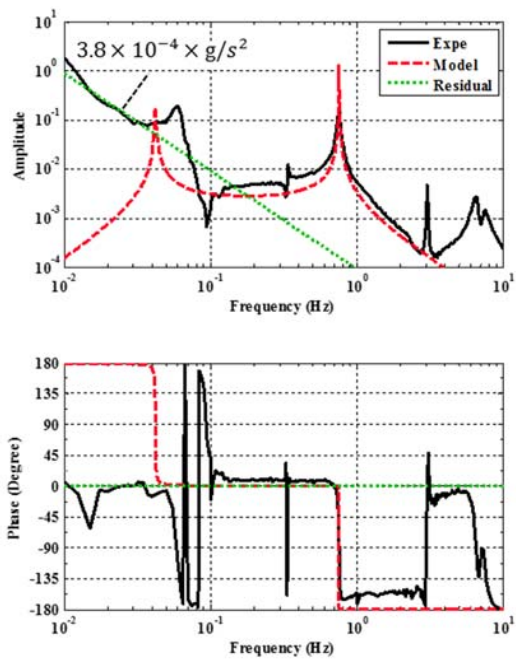

**Fig. 11.** Plot similar to Fig. 10, with the suspension tilt frequency tuned to 45 mHz instead of 150 mHz.

## 4 Translational response experiment

One of the main drawbacks of the suspended seismometer concept is that it reduces the translation sensitivity below the tilt frequency and above the pendulum frequency.

The goal of the experiment presented in this section is to verify that the signal induced by translation is in agreement with the model prediction. The test setup is configured as shown in Fig. 12. The rotation point is located well below the suspension point (about 1 meter), so that the driven force ($F$) creates large translation at the suspension point. In this configuration, the translation injected (~ 1m/rad) significantly exceeds the residual tilt coupling characterized in the previous section (~0.005 m/rad between the resonances), so that the translation component is the dominant input.

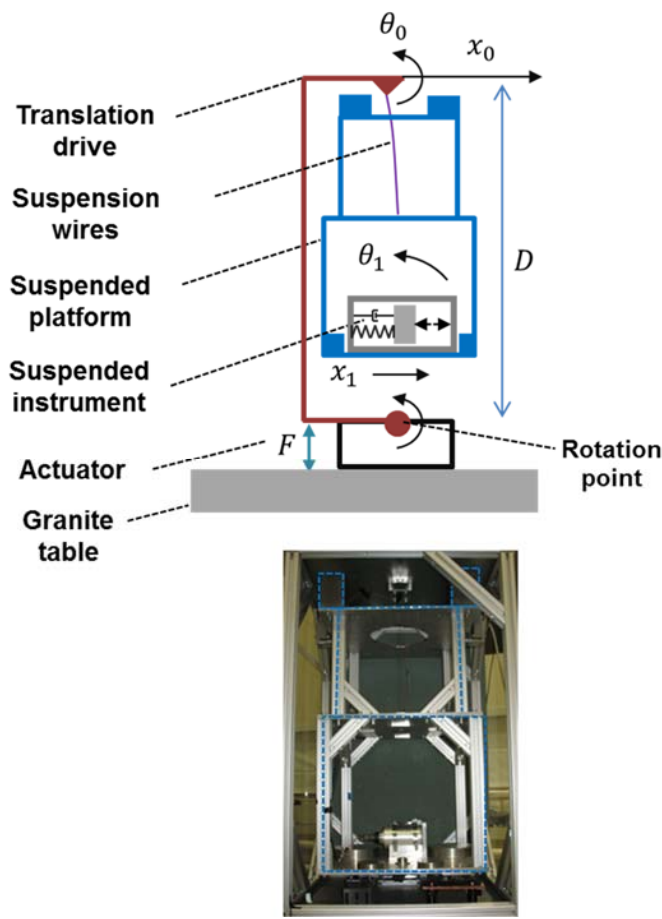

**Fig. 12.** Schematic and a picture of the experiment designed to inject a large ratio of translation over tilt.

The curves in Fig. 13 show the response of the suspended seismometer for four different tests using tilt frequencies ranging from 55 mHz to 190 mHz. As predicted by the models, the suspended seismometer acts as a band-pass filter, with sensitivity values near unity between the natural frequencies.

The response of the suspended seismometer tuned for a 155 mHz tilt frequency is compared with the response of the model in Fig. 14. The unwanted residual tilt coupling is about $3.95 \times 10^{-4}$ g/$s^2$. This value is higher than what was measured in the previous test where only tilt was injected.

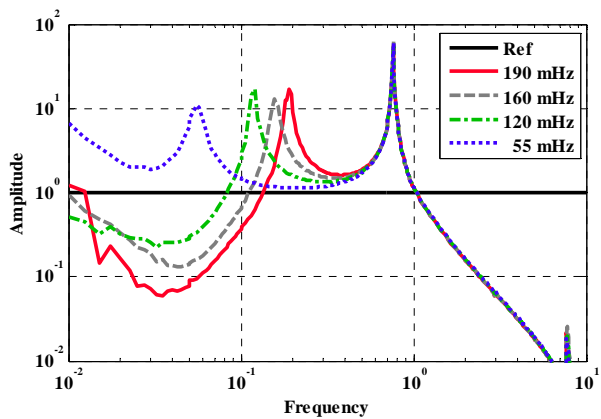

**Fig. 13. Experimental measurement of the translation sensitivity ($\widehat{\delta}_1/x_0$, black solid line) for different values of tilt frequency.**

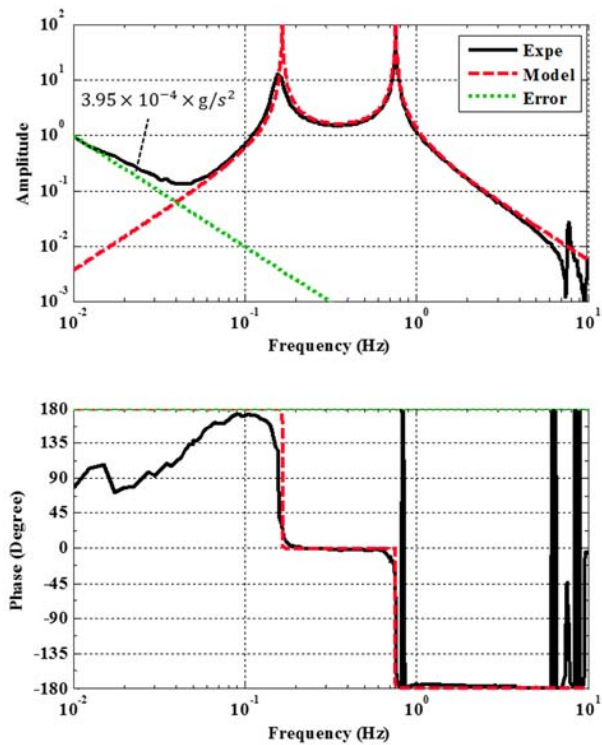

**Fig. 14. Transfer functions for a translation drive ($\widehat{\delta}_1/x_0$). Comparison of experimental results ("Expe" in the legend) and model. The difference (residual coupling) is fitted with the dotted line.**

Further experiments must be conducted to explain and reduce the sources of cross-couplings, but in this configuration, the low frequency cross coupling remains 2500 times lower than what would be sensed by a regular non-suspended seismometer.

A similar plot is shown in Fig. 15, where the tilt frequency has been reduced to 55 mHz. The residual coupling is significantly higher than in the previous measurement, with a value $3.2 \times 10^{-3}$ g/$s^2$ m/rad. These series of measurements show that the lower the tilt frequency, the higher the unwanted low frequency cross coupling. In the lowest frequency configuration the residual coupling is still 300 times lower than g/$s^2$.

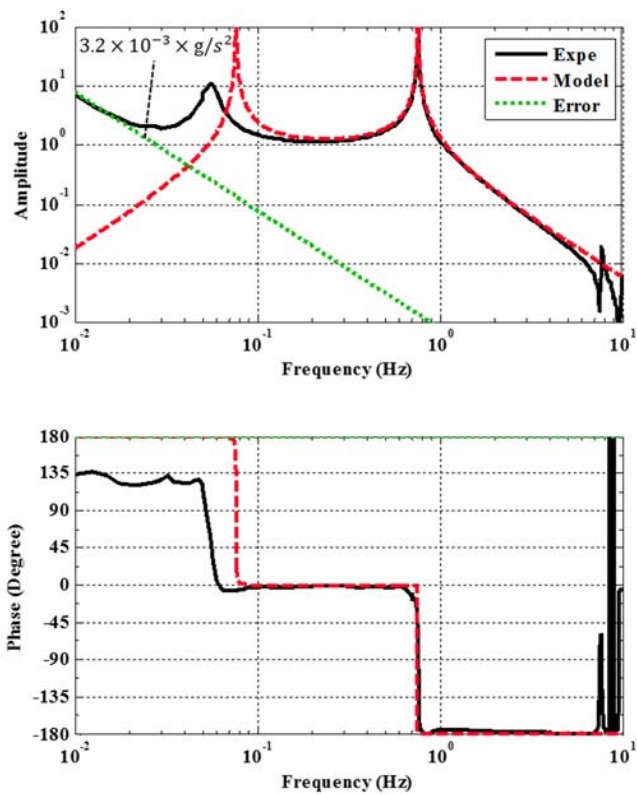

**Fig. 15. Plot similar to Fig. 14, with the suspension tilt frequency tuned to 55 mHz instead of 155 mHz.**

These experimental results presented in sections 3 and 4 demonstrated the effectiveness of this suspended seismometer approach. The next section summarizes the lessons learned during the course of development of this research, and discusses the prospects and applications.

## 5 Prototyping lessons and Prospects

While the previous sections presented the results obtained with our latest version of the suspension design, we find it important to highlight useful findings and lessons learned during the earlier development phases of this research. This section summarizes this information, and presents the prospects enabled by the results of this research.

## 5.1 Suspension configurations

There are many diffident technologies which can be used to suspend the seismometer. The joints of the suspension can be engineered with knife edges [18], flexures [27, 28], ribbons, metal wires, or silica fibers [29].

The first prototype built was a single wire suspension, shown in Fig. 16, which has the advantage of filtering the tilt in two directions. Unfortunately, such a configuration also features a low frequency torsion mode around the vertical axis which makes the system rather difficult to operate due to the very long settling time of this mode. Eddy current dampers were added to damp this mode but balancing the system along two axis remained quite challenging.

Several solutions were considered to raise the frequency of the torsion mode. Among the solutions tested, are the double wire configuration and the use of cross flexes. The cross flexes we designed, shown in Fig. 17, introduced significant cross couplings. We were not able to align them with sufficient accuracy to obtain satisfactory tilt filtering.

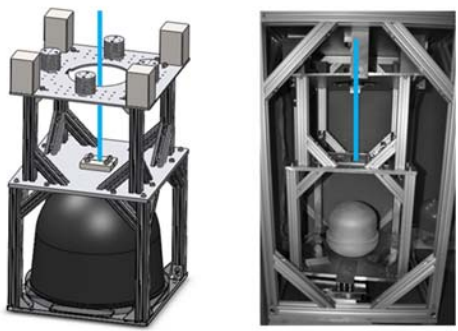

**Fig. 16.** Single wire suspension.

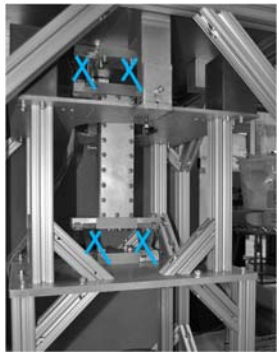

**Fig. 17.** Suspension using cross flexes.

The double wires solution shown in Fig. 18 proved to be practical, easy to balance and to commission. This is the configuration that was used for all the test results presented in the previous sections.

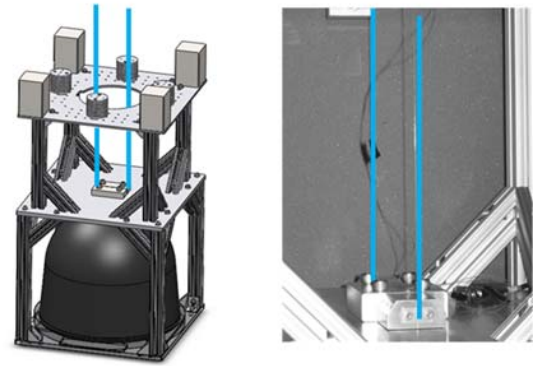

**Fig. 18.** Prototype with double metal wire.

The double wire solution can also be used to design a compact suspension in which the sensor is placed in between the wires as shown in Fig. 19.

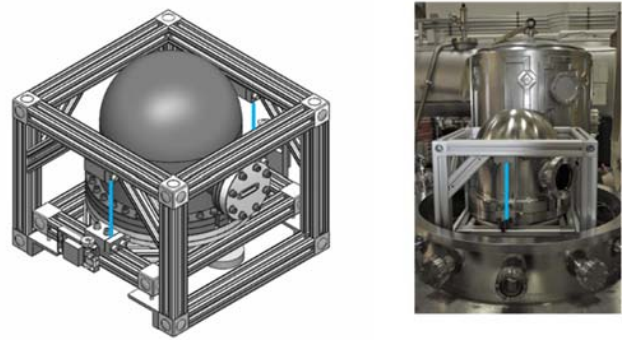

**Fig. 19.** Compact vacuum compatible version

## 5.2 Electrical wiring

During the testing phase it was found that the electrical wiring was one of the main limiting factors in the performance of the filter. For the tests performed with the passive seismometers, the best results were obtained when only two thin electrical wires were routed in parallel to the suspension wire to carry the signal to a pre-amp located beyond the suspension point. For the broadband seismometers, we replaced the heavy cable supplied by the manufacturer with light gauge wire as shown in Fig. 20. The influence of these cables on the tilt transmission remains to be quantified.

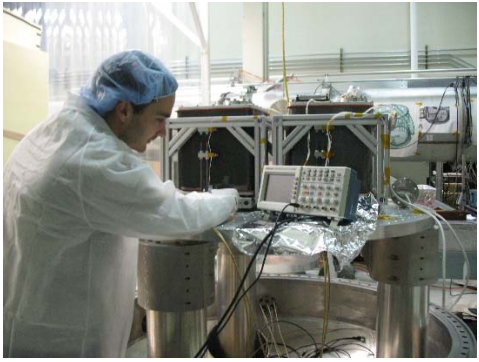

**Fig. 20. Custom electric cables routed along the suspension wires.**

### 5.3 Air currents

Sensitive systems such as the LIGO active isolation platforms must operate in vacuum to provide low-noise performance. For the driven tests presented in this article, a Styrofoam enclosure proved to be adequate shielding from air currents.

For low noise operation, the suspended instrument will need to operate in vacuum. Sealed pods can be used to encapsulate the broadband seismometers. Fig. 21 shows low cost pods using steel square tubing which can be used for prototyping phase. For sensitive vacuum systems, a more reliable pod design as shown in Fig. 19 can be used. Those are the pods designed for Advanced LIGO. These are filled with Neon used as a tracer of potential leaks to be detected by residual gas analysis. Additionally, the pods are equipped with a pressure sensor to help in identifying a potential pod leak.

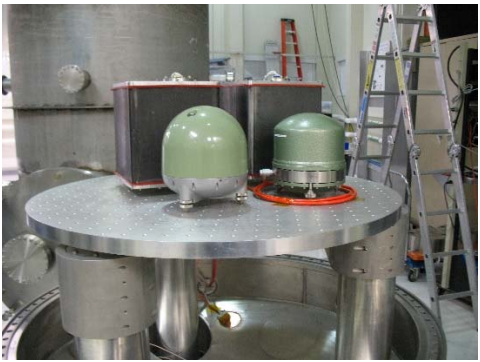

**Fig. 21. Seismometers pods for prototyping phases.**

Tests performed in vacuum with the suspensions showed that the suspensions were drifting for several hours after the chamber was closed and evacuated. The transient behavior is shown in Fig. 22. The curves show the measurement of the angle of the platform performed with capacitive position sensors measuring the differential motion between the suspended platform and the support table.

The blue and the green curves show the angular motion the two suspensions. At the beginning of the measurement, the two suspensions are similarly excited by a large transient excitation (the closing of the chamber). In the following minutes, the time series show an exponential decay induced by the dissipation in the joints. The suspensions then drift for several hours before settling. This behavior must be further investigated to identify the cause and estimate the possible consequences on the noise performance. One possible explanation is related to the power dissipation of the electronics of the seismometers, which require long periods to reach thermal equilibrium.

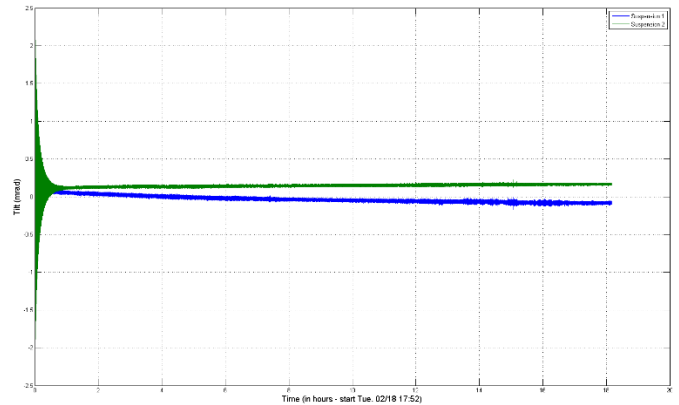

**Fig. 22. Time series of the suspension tilt.**

### 5.4 Broadband seismometer test

Passive seismometers (geophones) were used for the prototyping phases and for the driven tests presented in the previous sections, as their robustness and short settling time were convenient for these phases of the project.

Broadband seismometers are, however, necessary for low noise operations such as those required for the seismic isolation of gravitational wave detectors. After the suspension had been developed and tested with the geophones, it was equipped with broadband seismometers (Trilium T240) as shown in Fig. 23, and the platform was re-tuned to obtain desired tilt frequencies.

The curves in Fig. 24 show the results obtained with the suspended broadband seismometer. The results are similar to those obtained with the geophones. One noticeable difference is the amplitude at the resonance which is higher than with the geophone. Modeling and simulations including the coupling between the moving mass of the geophone and the suspended platform show that the internal damping of the geophone tends to dampen the resonance of the suspended platform. For the next generation of suspended seismometers, the amplitude at the tilt resonance will need to be addressed adequately to ensure robust operations.

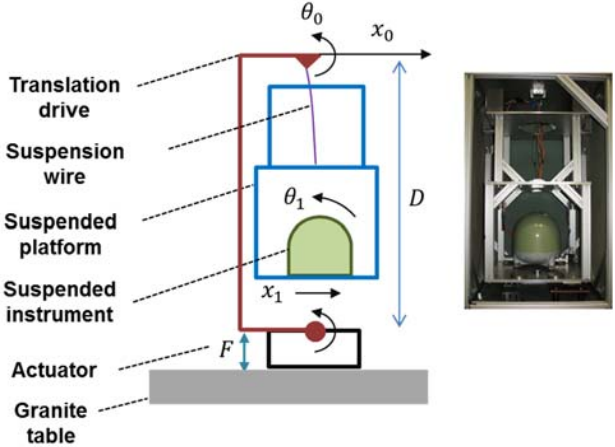

**Fig. 23.** Suspended seismometer test SETUP.

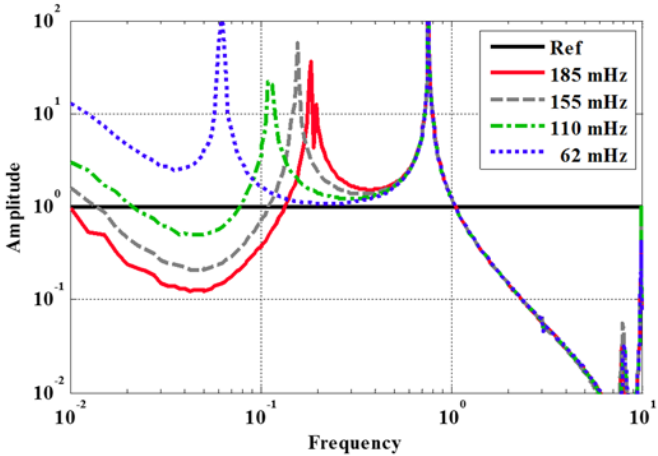

**Fig. 24.** Transfer functions for a translation drive ($\hat{\delta}_1/x_0$) using a broadband seismometer at different tilt frequency tunings.

## 6 Conclusion and Prospects

The goal of this research was to investigate the use of suspensions to reduce the transmission of tilt motion to horizontal inertial sensors.

The modeling of the suspended seismometer concept showed that an ideal suspension can attenuate the tilt transmission by orders of magnitude, while maintaining adequate translation sensitivity and measurement noise in the bandwidth of interest.

The experimental results presented showed that the response of the suspended seismometer is in good agreement with simulation results. The un-modelled residual tilt signal due to tilt coupling is orders of magnitude lower than the tilt signal in a non-suspended seismometer.

The phases of prototyping and testing showed that the two-wire suspension is a practical solution. The sources of cross-coupling and residual tilt-transmission should be investigated to further improve the filtering performance. Internal damping solutions to provide adequate dynamic range must be studied with careful attention to thermal noise [30]. In-vacuum low-noise laboratory tests should be conducted at a quiet site, where environmental conditions are more similar to those of a gravitational wave detector sites.

The results of this investigation indicate that the suspended seismometer approach is a viable solution to improve the low frequency seismic isolation performance of gravitational wave detectors.

A suspended seismometer could be aligned with the suspension point of an Advanced LIGO quadruple suspension [29], attached to the two-stage isolation system [31, 32]. The signal of this instrument could be used in a sensor fusion scheme to drive the supporting isolation platform and reduce the longitudinal motion in the 4km optical cavities.

### Acknowledgements

This work was carried out within the LIGO laboratory. LIGO was constructed by the California Institute of Technology and Massachusetts Institute of Technology with funding from the National Science Foundation, and operates under cooperative agreement PHY-0757058. Advanced LIGO was built under award PHY-0823459. This document has been assigned LIGO Laboratory document number LIGO-P1400061. This work would not have been possible without the outstanding support of the LIGO visitor program. The authors are very grateful to Brian Lantz for encouraging us and our community in finding solutions to mitigate tilt noise in active isolation systems, to Dan DeBra for his very useful advices on flexure mechanisms, and to Krishna Venkateswara and Vladimir Dergachev for many useful discussions on tilt sensing problems and instrumentation.